\documentclass[aps,amsmath,amssymb,prb,preprint]{revtex4-1}
\usepackage{amsmath, latexsym, color, graphicx, tabularx, booktabs, enumerate}
\usepackage{multirow}
\usepackage[normalem]{ulem} 
\usepackage{caption}
\usepackage{epstopdf}
\usepackage{mathtools}
\usepackage{hyperref}
\usepackage{cleveref}

\begin{document}
\title{Opening band gaps of low-dimensional materials at the meta-GGA level of density functional approximations}
\author{Bimal Neupane}
\email{bimal.neupane@temple.edu}
\author{Hong Tang }
\email{hongtang@temple.edu}
\author{Niraj K. Nepal}
\author{Santosh Adhikari}
\author{Adrienn Ruzsinszky}

\affiliation{Department of Physics, Temple University, Philadelphia, Pennsylvania 19122, United States}

\begin{abstract}
The quasiparticle band structure can be properly described by the GW approximation, at an increased computational cost. Semilocal density functionals up to the generalized gradient approximation (GGA) level cannot compete with the accuracy of hybrid-based approximations or GW. Meta-GGA density functionals with a strong dependence on the kinetic energy density ingredient can potentially give wider band gaps compared to GGA's. The recent TASK meta-GGA density functional [Phys. Rev. Research, \textbf{1}, 033082 (2019)] is constructed with an enhanced nonlocality in the generalized Kohn-Sham scheme, and therefore harbors great opportunities for band gap prediction. Although this approximation was found to yield excellent band gaps of bulk solids, this accuracy cannot be straightforwardly transferred to low-dimensional materials. The reduced screening of these materials results in larger band gaps compared to their bulk counterparts, as an additional barrier to overcome. In this work we demonstrate how the alteration of exact physical constraints in this functional affects the band gaps of monolayers and nanoribbons, and present accurate band gaps competing with the HSE06 approximation. In order to achieve this goal, we have modified the TASK functional (a) by changing the tight upper-bound for one or two-electron systems ($h_X^0$) from 1.174 to 1.29 (b) by changing the limit of interpolation function $f_X (\alpha \rightarrow \infty$) of the TASK functional that interpolates the exchange enhancement factor $F_X (s,\alpha)$ from $\alpha=$ 0 to 1. The resulting modified TASK (mTASK) was tested for various materials from 3D to 2D to 1D (nanoribbons), and was compared with the results of the higher-level hybrid functional HSE06 or with the G$_0$W$_0$ approximation within many-body perturbation theory. We find that mTASK greatly improves the band gaps and band structures of 2D and 1D systems, without significantly affecting the accuracy of the original TASK for the bulk 3D materials, when compared to the PBE-GGA and SCAN meta-GGA. We further demonstrate the applicability of mTASK by assessing the band structures of TMD nanoribbons with respect to various bending curvatures.
\end{abstract}
\maketitle

\section{Introduction}
Due to their reduced dimensionality, two-dimensional (2D) materials (e.g., transition metal dichalcogenides or TMDs) exhibit an extraordinary optical response \cite{RAH14, UBS14} in comparison with bulk counterparts, as has been shown early via the examples of graphene and 2D MoS$_{2}$.  The spatial confinement and reduced dielectric screening of 2D materials causes strong Coulomb interaction that allows more stable exciton formation with large binding energy and oscillator strength compared to bulk crystals. These attractive features can in principle be harvested for optoelectronics. Fine tuning of the optical properties can be achieved by mechanical bending that alters the electronic structure. Optical nanodevice functionality can require the optimization of exciton binding energy, optical absorption and most importantly the fundamental band gap with respect to the strains existing in the bending space. Strain engineering \cite{CRC13} has been known as a tool to control the electronic properties of 2D materials, but the impact of bending is not yet thoroughly explored.  

Fundamental band gaps carry a great relevance from a theoretical and practical view \cite{PYB17}. Since the fundamental band gap is the unbound limit of an exciton series, it controls the optoelectronic response of materials. Optical absorption spectra within time dependent DFT (TDDFT) are usually evaluated on a scissor-operator-corrected DFT band gap \cite{LA89,DG93}. This approach is simpler than applying the more expensive GW approximation, but the scissor-shift correction is often evaluated from experiments, and the correction is very unlikely to be available for strained or bent structures relevant for industrial applications. In bulk crystals, GW yields a nearly constant shift to the fundamental gap of semilocal DFT even for strained structures, but this is not necessarily true any more for low-dimensional systems \cite{SL13, WSS19}. This difference is a consequence of enhanced many-body effects present at low-dimensionality. The literature reports both enhanced and weaker band gap change with GW compared to Kohn-Sham DFT, for the fundamental and higher band gaps of stretched or compressed carbon nanotubes \cite{LA89}. A similar effect can be expected for nanoribbons. 
As an alternative to the missing scissor corrections for the bent structures, we can use the hybrid functional HSE06 \cite{HSE03,HSE03Erratum} in a generalized Kohn-Sham scheme to estimate the band gaps and bandstructures for various bending curvatures. HSE06 is known to yield reasonably accurate band gaps for bulk crystals. Less information is available about the accuracy of band gaps of nanoribbons with HSE06, and the increased size of supercells can make such calculations more demanding. The demand for an alternative approximation with more computational feasibility is large, but semilocal density functionals are far from being reliable for accurate band gap prediction. The underestimation of band gap in semilocal functionals is a consequence of the lack of derivative discontinuity within the Kohn-Sham potential \cite{PPL82,PL83,SS83,P85}. Accurate band gaps require an effective potential that is either a discontinuous multiplicative operator or a continuous nonlocal operator. Early band-structure calculations showed that the LSDA band gaps for semiconductors were often about half the measured fundamental energy gaps.
	Meta-GGA density functional approximatons are placed on the third rung of density functional approximations \cite{PRT05}. Meta-GGA's are explicit functionals of the Kohn-Sham orbitals and implicit functionals of the density. Some meta-GGA's were already found to open band gaps more than GGA's \cite{SRP2015}. The successful SCAN \cite{SRP2015} meta-GGA usually yields a slight improvement for bulks solids, but it is still far from approaching the accuracy of HSE06 or GW. Alternatively the recently developed TASK meta-GGA \cite{AK19} for band gaps of bulk solids in a generalized \cite{PYB17,SGV96} Kohn-Sham scheme is a promising approximation. The TASK meta-GGA is not designed to yield comparable accuracy for ground-state properties; capturing accurate ground-state properties and accurate electronic structures simultaneously is not to be expected from the first three rungs of Jacob's ladder of density functional approximations. Nevertheless, our focus in this work is on band gaps and not on ground-state properties.
TASK and similar meta-GGA’s, implemented within the generalized Kohn-Sham scheme, could provide improved estimates of quasiparticle energies. Further hints refer to the TASK meta-GGA as a potential tool for excitonic peaks in the spirit of Nazarov and Vignale \cite{NV11}, taking the second functional derivative of the exchange-correlation energy. With the consideration of meta-GGA’s for band gaps we follow an approach like that of the Tran-Blaha potential functional \cite{TB09}, with the change that the potential is properly a functional derivative of the exchange-correlation energy. While the TASK functional has been tested on a set of bulk solids with great accuracy as a result, its performance is completely unknown for low-dimensional materials. By now, tremendous numbers of references point out the enhanced optical response from the different screening in low-dimensional materials compared to bulk solids \cite{UBS14,QFL13,R12,HKZ14}.

\section{Methods}
Typical meta-GGA’s utilize the kinetic energy density as an additional ingredient beyond the ones of GGA’s \cite{PRT05,SRP2015,AK19}. The kinetic energy density makes the meta-GGA form an implicit density functional, in principle, with potential \cite{PYB17} for band gap prediction and optical spectra.  Accurate band gap prediction is based on the derivative discontinuity \cite{PPL82}; a feature that semilocal density functional approximations are missing. The derivative discontinuity manifests itself as a step structure in the exact exchange-correlation potential as \cite{kummel2003simple}
\begin{equation}
    \Delta_X={v_X\left(\textbf{r}\right)|}_+-{v_X\left(\textbf{r}\right)|}_-{ = \frac{{\delta E}_X\left[n\right]}{\delta n\left(\textbf{r}\right)}|}_+-{\frac{{\delta E}_X\left[n\right]}{\delta n\left(\textbf{r}\right)}|}_{-,}                                                 
\end{equation}
\noindent where the positive and negative signs refer to the left and right-hand sides of the potential at integer electron numbers. Perdew and Levy showed that the functional derivative discontinuity of the exchange-correlation density functional plays a crucial role in the correct prediction of band gaps \cite{PL83}.
 The functional derivative of a meta-GGA within the Kohn-Sham scheme contains the $\tau$-dependence in the third term below that increases the derivative discontinuity \cite{EH14}
 \begin{equation}
  \frac{{\delta E}_X^{mGGA}\left[n\right]}{\delta n\left(\textbf{r}\right)}=\frac{\delta}{\delta n\left(\textbf{r}\right)}\int{e_X\left(n,\nabla n,\tau\right)}d^3r^\prime=\frac{\partial e_X}{\partial n}\left(\textbf{r}\right)-\nabla\cdot\left[\frac{\partial e_X}{\partial\nabla n}\left(\textbf{r}\right)\right]+\int{\frac{\partial e_X}{\partial\tau}\left(\textbf{r}^\prime\right)\frac{\partial\tau\left(\textbf{r}^\prime\right)}{\partial n\left(\textbf{r}\right)}}d^3r^\prime
 \end{equation}
 
 To date, only a few meta-GGAs have been identified to display the necessary dependence on the kinetic energy density \cite{SRP2015}. A recent effort \cite{AK19} demonstrates that meta-GGA’s with an enhanced dependence on the dimensionless kinetic energy density ingredient $\alpha=\frac{\left(\tau-\tau^w\right)}{\tau^{unif}}$, give accurate band gaps and can be applied to optical properties. Band gap calculations with the TASK meta-GGA have been done within the generalized Kohn-Sham scheme, in which the exchange-correlation potential is a differential operator with significant nonlocality \cite{AK03}. The dimensionless iso-orbital indicator $\alpha$ contains $\tau^w$, the kinetic energy of a one-electron system and $\tau^{unif}$,the kinetic energy density of the homogeneous electron gas. The condition that is responsible for an improved band gap as well as for excitonic peaks is: 
 \begin{equation}
       \frac{\partial e_X}{\partial\tau}>0
 \end{equation}
 \noindent $\frac{\partial e_X}{\partial \tau}$ is a $\tau$-dependent factor that contributes to a first or second functional derivative of the exchange-correlation potential or the corresponding energy.
For practical use it is easier to show the $\tau$-dependence of the exchange enhancement factor $F_X$. A sizeable nonlocality leads to the condition \cite{NV11}:
\begin{equation}
    \frac{\partial F_X}{\partial\alpha}\ll 0. 
\end{equation}
\noindent The recent SCAN meta-GGA was found to exhibit an observable opening of band gaps of various bulk solids. Still, SCAN is a ground-state density functional. 
The TASK meta-GGA approximation is a thoughtful re-construction of the exchange form of SCAN so that with the increased slope $\frac{\partial F_X}{\partial\alpha}$ more nonlocality in the exchange potential can be achieved. The TASK exchange basically keeps the exchange $F_X$ of SCAN
\begin{equation}
    F_X^{TASK}\left(s,\ \alpha\right)=h_X^0g_X\left(s\right)+\left[1-f_X(\alpha)\right]\left[h_X^1\left(s\right)-h_X^0\right]\left[g_X\left(s\right)\right]^d
\end{equation}
\noindent In the expression of $F_X$, $h_X^0$ is the tight upper-bound for one or two-electron systems. The same $h_X^0$ was applied in the SCAN functional. For $\alpha=$0, $F_X^{TASK}\le1.174$, which is also the conjectured bound for this limit \cite{PRS14}. $f_{X}(\alpha)$ is a function that satisfies the fourth-order gradient limit and interpolates between $\alpha$=0 and $\alpha=$1. When $f_X\left(\alpha=0\right)=$1, the TASK functional recovers SCAN by satisfying the two-dimensional scaling \cite{PRS14,L91,PP00} of the reduced gradient $s=\frac{\left|\nabla n\right|}{2\left(3\pi^2\right)^{1/3}n^{4/3}}$ as,

\begin{equation}
    {\ \ g}_X\left(s\right)=1-e^{\left(-c s^\frac{1}{2}\right)}.                                         
\end{equation}

\noindent In equation (6), c is a parameter that was found by fitting to the exact atomic energy of hydrogen. TASK deviates from SCAN in the construction of $h_X^1\left(s\right)$. The construction of $h_X^1\left(s\right)$ and $f_X(\alpha)$ aims to increase the slope defined by equation (4). The fourth-order gradient expansion around $s=$0 and $\alpha=$1 recovered by both $h_X^1\left(s\right)$ and $f_X(\alpha)$ together is

\begin{equation}
    {F}_X^{GE4}\left(s,\alpha\right) \sim 1+\mu_ss^2+\mu_\alpha\left(\alpha-1\right)+C_ss^4+C_{s\alpha}s^2\left(\alpha-1\right)+C_\alpha\left(\alpha-1\right)^2+\mathcal{O}\left(\nabla\right)^6
\end{equation} 

\noindent
along with a fourth-order expansion for $\tau$ \cite{BJC76,PSH86}, $\mu_s$, $\mu_\alpha$, $C_s$, and $C_{s\alpha}$ are chosen to correspond to the coefficients of the Taylor expansions for $h_X^1\left(s\right)$ and $f_X\left(\alpha\right)$ to fourth-order, respectively. All these coefficients and the exact conditions for $f_X\left(\alpha\right)$ lead to eight equations for the eight coefficients of Chebyshev expansions \cite{B87,PTV92} for $h_X^1\left(s\right)$ and $f_X\left(\alpha\right)$ to fourth-order.  $\mu_\alpha$ is determined from a quadratic equation with the ingredient of $h_X^0$. The more negative value of $\mu_\alpha$ yields an increased nonlocality compared to any previous meta-GGA’s, so that TASK delivers excellent fundamental band gaps for bulk solids with a large enough variety of structures.\\

In the last decade, utilizing methodological and computational advances \cite{G09}, hybrid functionals \cite{BB00} have been increasingly used to investigate a variety of periodic systems with plane wave basis sets. Among hybrid functionals the HSE06 approximation \cite{HSE03,HSE03Erratum} is particularly popular, but some other hybrid approximations \cite{YTH04,WHS09,AB99,SGG14,RJS15} with various ratios of exact exchange admixture have also become beneficial for the condensed matter community. The most successful hybridization schemes are based on partitioning the Coulomb operator into short- and long-range components in a two-parameter form in which the parameters $\alpha$ and $\beta$ control the mixing ratio of long- and short-range exchange. This scheme is the basis of various popular density functionals, such as CAM-B3LYP \cite{YTH04}, LC-$\omega$PBE \cite{WHS09}. HSE06 with $\beta =$  0.25 and $\omega = $ 0.11 Bohr$^{-1}$ is a short-range screened hybrid that recovers PBE0 for the short range therefore enabling computational efficiency for periodic systems.
The mixing of exact exchange in hybrid functionals is obviously linked to spatial nonlocality. The hybrid parameter $\alpha$ is often treated as an adjustable parameter to reproduce the experimental band gap of solids \cite{PH11,C12,ABD08,ABP11,BAP10}. Since in semiconductors and insulators the screening of the long-range tail of the Coulomb interaction was found to be proportional to the inverse of the static dielectric constant ${(\epsilon}_\infty^{-1})$, it has become natural to link $\alpha$  to $\epsilon_\infty^{-1}$. It should be also noted that an analogy was established between hybrid functionals and the static COulomb Hole plus Screened EXchange (COHSEX) approximation \cite{H65}, in which the screened Coulomb interaction is expressed as
\begin{equation}
    W(\mathbf{r},\mathbf{r}^\prime)\approx\frac{1}{\epsilon_\infty}v(\mathbf{r},\mathbf{r}^\prime) 
\end{equation}

\noindent Our work, although in a completely different way, seeks the same spatial nonlocality needed for band gaps, considering the reduced screening with increasing low-dimensionality. Semilocal density functionals such as GGA’s  do not exhibit spatial nonlocality, but through $\alpha$ meta-GGA’s \cite{BE90,SJF92,SNW97}  are implicit functionals of the density, and can overcome this limitation \cite{SRP2015,SXR12,SXF13} . The dimensionless ingredient $\alpha$  in the TASK and SCAN functional is an explicit functional of the Kohn-Sham orbitals, and therefore carries potential spatial nonlocality. Note that the notational coincidence with the hybrid functionals’ $\alpha$, traditionally called, is originally accidental, but now gains some physical analogy. 
Semiconductors have a significantly lesser screening than metals, as a consequence of their less-localized exchange-correlation hole. Meta-GGA functionals based on the Laplacian of the electron density and not on the kinetic energy density can better describe the more localized exchange-correlation hole of metals than the one of semiconductors \cite{MT18,MT20,mejia2017deorbitalization,KaplanAPS}. Meta-GGA's with gradient and Laplacian-only ingredients were found more accurate for metals than for semi-conductors, while SCAN and other meta-GGA's with orbitals in their $\alpha$ or $\tau$-dependence work better for semiconductors and insulators. With a proper modification of the original TASK functional the slope $\frac{\partial F_X}{\partial\alpha}$ can be further increased so that the modified TASK (mTASK) has screening appropriate for low-dimensional materials.

In Section 4 we will discuss how the increased nonlocality with the corresponding increased screening affects 3D, 2D and 1D materials, when the fundamental band gaps of these materials with various dimensionality are compared to HSE06. In our modification (modified TASK or mTASK) we focus on the coefficient $\mu_\alpha$. An increased slope is equivalent to increased $\alpha$-dependence of $F_X$.
Based on the fourth-order gradient expansion of $F_X$, $\mu_\alpha$ is determined from a quadratic equation with $h_X^0$ as an ingredient 
\begin{equation}
    \mu_\alpha^\pm=-\frac{97+3h_X^0\pm\sqrt{9\left(h_X^0\right)^2+74166h_X^0-64175}}{1200}. 
\end{equation}

\begin{figure}[h!]
    \centering
    \includegraphics[scale=0.4]{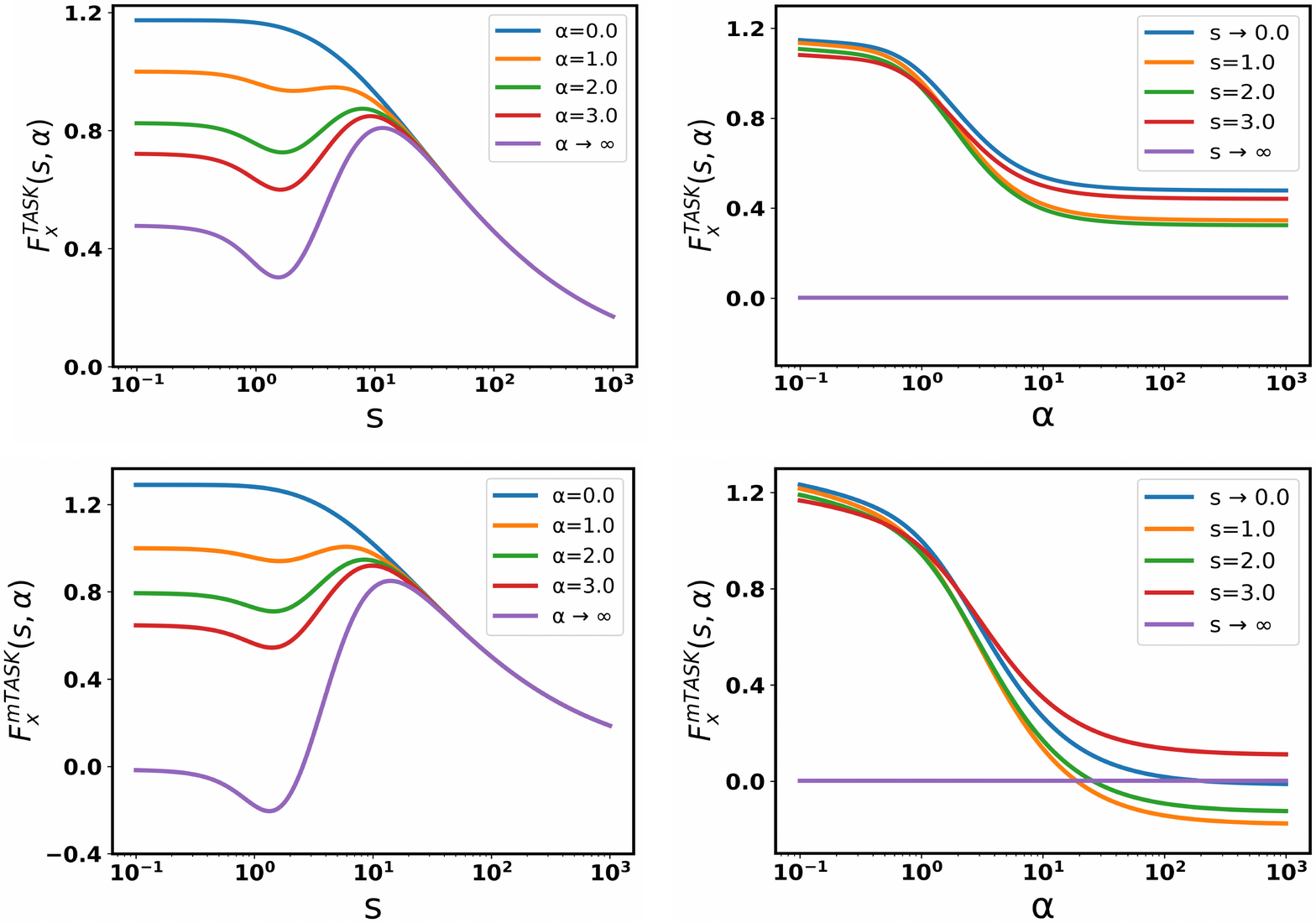}
    \caption{The enhancement factors $F_X$ for the TASK (upper panel) and modified TASK (lower panel) as functions of s and $\alpha$, respectively.   }
    \label{fig:fig1}
\end{figure}

\noindent This TASK expression is able to deliver a large-enough nonlocality $\mu_\alpha^+=-0.209897$ for bulk solids, but not for low-dimensional materials. We realize that the increased band gap from the reduced screening in monolayers and nanoribbons requires a different   $\mu_\alpha^+$  and therefore different $h_X^0$. Lacking more exact constraints for the ground state, with any choice for ${F_X \le h}_X^0 \neq 1.174$, we have to sacrifice the tight Lieb-Oxford bound. In this work, we choose $F_X\le h_X^0=1.29$. This choice is far from the Lieb-Oxford bound $F_X\le1.804$  known for GGA’s. This condition is explained in Ref. \cite{PRS14} for meta-GGA functionals. To make the exchange enhancement factor a smooth and monotonically decreasing function of $\alpha$ for any value of s, we choose the limiting condition $f_X\left(\alpha\rightarrow\ \infty\right)=\ -3.5$. From Figure \ref{fig:fig1}, it can be seen that $\frac{\partial F^{mTASK}_X}{\partial\alpha}\le\frac{\partial F^{TASK}_X}{\partial\alpha}$ for any value of s. We expect the inclusion of more nonlocality in the mTASK exchange-correlation potential compared to TASK provides better band gaps and band structures for 2D materials and nanoribbons. In addition, when choosing the value of $h_X^0=1.804$ known for GGA’s, we have observed that the exchange enhancement factor is not a smooth function of $\alpha$.\\

\begin{figure}[h!]
    \centering
    \includegraphics[scale=0.4]{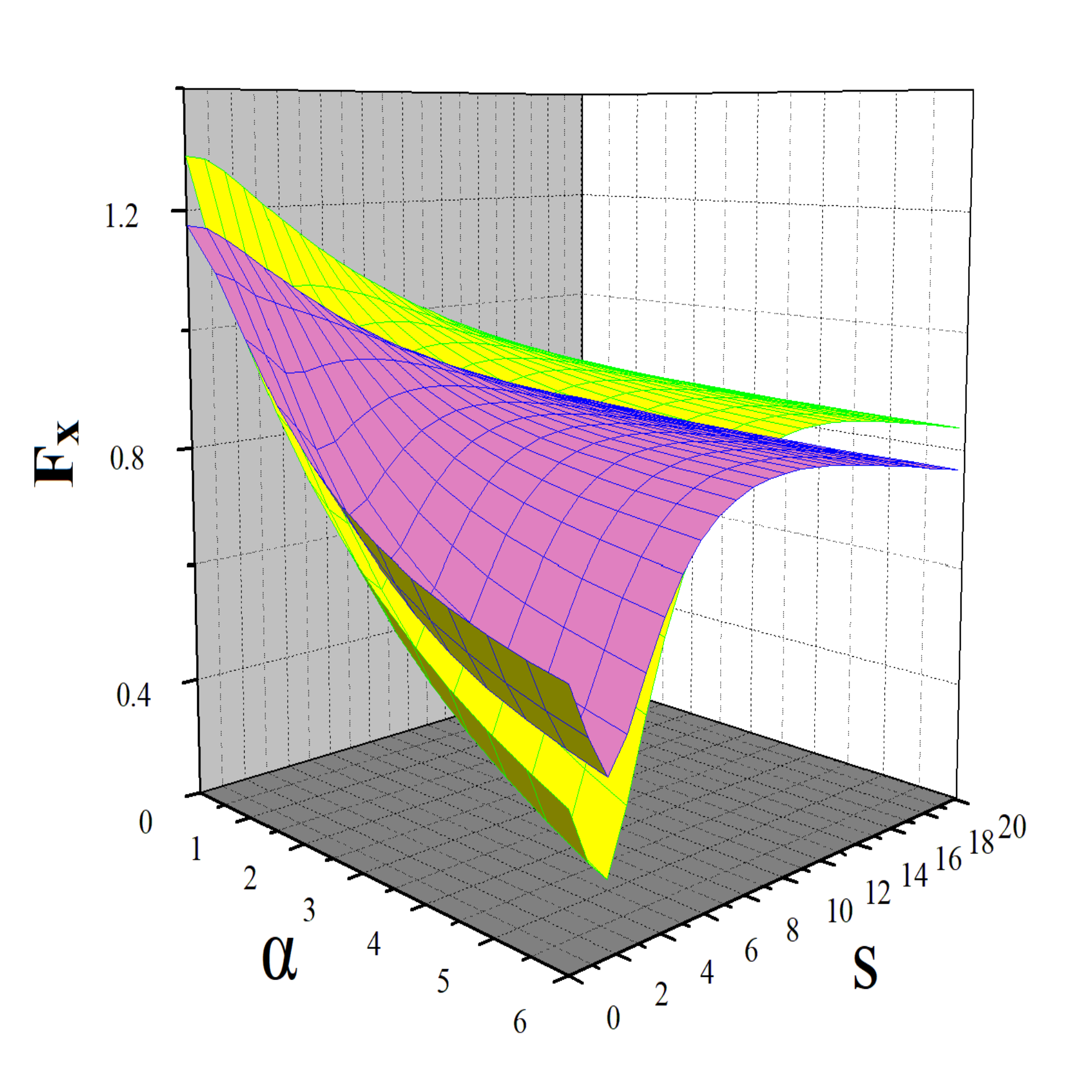}
    \caption{The three-dimensional contour plot of the exchange enhancement factor $F_X$ as a function of  $\alpha$ and s for the original TASK (purple, $h_X^0=$ 1.174, $\mu_\alpha=-$ 0.209897, $f_X\left(\alpha \rightarrow \infty \right)=-$ 3 ) and the modified TASK (yellow, $h_X^0=$ 1.29, $\mu_\alpha=-$ 0.231993, $f_X\left(\alpha\rightarrow\infty\right)=-$ 3.5). }
    \label{fig:fig2}
\end{figure}

Figures \ref{fig:fig1} and \ref{fig:fig2} demonstrate the changes in mTASK compared to TASK. Figure \ref{fig:fig1} presents $F_X$ for s and $\alpha$ respectively. $F_X$ of mTASK starts out at $h_X^0=$ 1.29, compared to $h_X^0=$ 1.174 of TASK. The larger $\alpha$- and s- dependence of mTASK is evident in Figure \ref{fig:fig1}.  The three-dimensional contour plot of $F_X$ (Figure \ref{fig:fig2}) allows showing all s and $\alpha$-dependence.

\section{Computational Details}
All calculations were performed in the Vienna \textit{ab initio} simulations package (VASP) \cite{kresse1996efficient,kresse1999ultrasoft}. The valence electrons of all elements, are treated by the projector augmented wave (PAW) pseudo-potential method \cite{blochl1994projector}, which is recommended in the VASP manual.  The pseudopotential for tungsten atom with valence electron configuration 6S$^{1}$5D$^5$ was utilized.  The plane-wave energy cut-off is set as 450 eV for all calculations and it leads to converged results. The Brillouin zone is sampled by a Gamma centered mesh of 8$\times$1$\times$1 for both the hexagonal armchair TMD nanoribbons and trigonal  TMD nanoribbons. The Gamma centered k-point mesh of 18$\times$18$\times$1 and 20$\times$20$\times$20 were used for TMD monolayers and bulk solids, respectively. 
A vacuum layer of more than 15 \AA \hspace{0.15cm}was inserted along both the width and thickness directions to prevent interactions between periodic images due to the long-range Coulomb interactions. The bent nanoribbons were constructed by fixing the distance between the two edge metal atoms (i.e M in MX$_2$). Two hydrogen atoms are attached to every edge metal atom and one hydrogen atom is attached to every edge X atom in the MX$_2$ nanoribbon. The edge atoms for the bent nanoribbons were only allowed to relax in the periodic direction. All other atoms were fully relaxed until the force on each atom became less than 0.01 eV/\AA. 
For the bulk solids we used experimental lattice constants in the calculations, in order to directly compare results with references. The PBE functional was used to optimize (or relax) the nanoribbon structures, and we utilized them to calculate the band structures using various DFT approximations. The lattice parameter along the periodic direction of the nanoribbon was also relaxed.

\section{Results}
\subsection{Bulk crystals}
\begin{figure}
    \centering
    \includegraphics[scale=0.4]{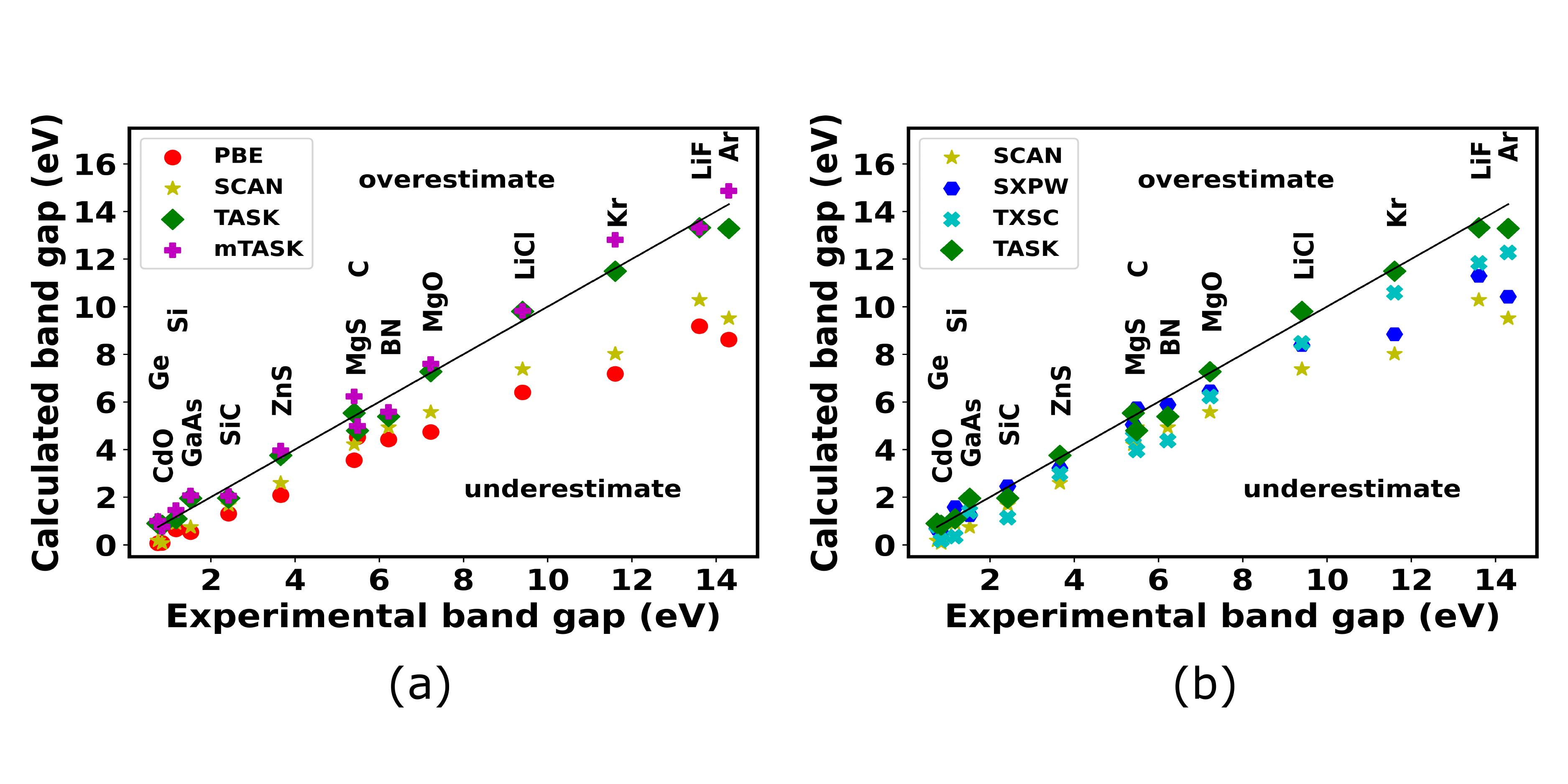}
    \caption{Comparison of calculated and experimental band gaps of bulk solids with various density functional approximations (a) PBE, SCAN, TASK, and mTASK (b) SCAN, SXPW, TXSC, and TASK. TXSC is a functional with the exchange component of TASK and the correlation from SCAN. SXPW refers to the exchange part of SCAN and correlation of PW92.}
    \label{fig:fig3}
\end{figure}

Bulks solids and low-dimensional materials often exhibit dramatically different physical properties, especially for optical response. To obtain a general picture about the applicability of the approximations considered in this work, first we have assessed several density functional approximations including TASK and mTASK for bulk solids. Figure \ref{fig:fig3} represents a correlation between the experimental and calculated band gaps of the same set of bulk solids from Ref \cite{AK19}.
The underestimation of band gaps from PBE is not surprising. As a semilocal density functional approximation, PBE has only the ingredients of the local density and the gradient of the local density, without explicit inclusion of nonlocal exchange effects, the latter being important for an accurate description of band gaps. The SCAN meta-GGA yields a systematic and slight improvement. As a meta-GGA functional, SCAN can include some nonlocal exchange effects through the orbital dependent ingredient $\alpha$. Although SCAN is largely accurate for equilibrium structures and energies for various bonds, the nonlocality derived from $\alpha$ may not be pronounced due to cancellation within the exchange and correlation parts. TASK proves to be very accurate, almost in all regions of band gaps except for the Ar crystal that has the largest band gap in this set and is underestimated by TASK. The enhanced screening in mTASK results in a slight overestimation of band gaps for bulk solids. \\

The effect of the correlation is exemplified through the SXPW and TXSC approximations. In the former  method SX refers to SCAN exchange, while the correlation component is from the PW92 local spin density approximation \cite{perdew1992accurate}. In the latter, TX is the TASK exchange, and SC is the correlation of SCAN.
SXPW and TXSC can be directly compared with SCAN and TASK, respectively, because they have  the same exchange but different correlation approximations. In general, both these approximations open the band gaps slightly more than SCAN, but they underestimate them compared to TASK. This can serve as an evidence for the cancellation effects in SCAN’s exchange and correlation parts for the nonlocality needed for band gap description. The improvement of the band gaps from TASK and mTASK is more consistent for the large-gap crystals such as MgO, LiCl, Kr, LiF, and Ar, as can be seen in Figure \ref{fig:fig3} a and b.

\subsection{Band gaps of single layers of 2D TMDs}
\begin{figure}
    \centering
    \includegraphics[scale=0.4]{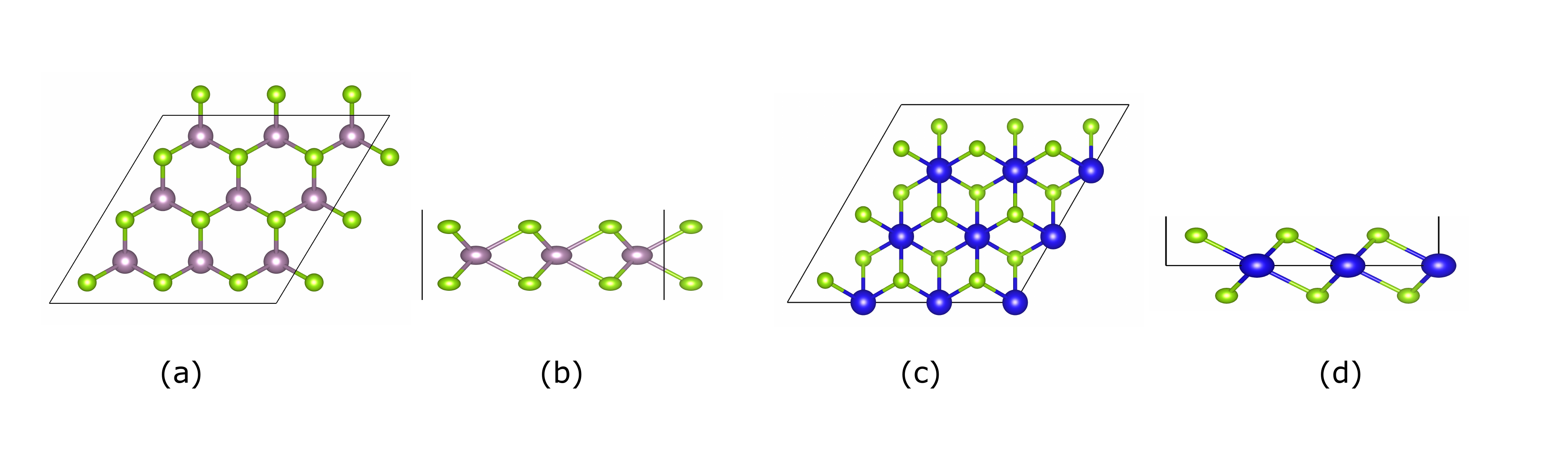}
    \caption{(a) Top view and (b) side view of monolayer 3$\times$3$\times$1 1H-TMDs. (c) Top view and (d) side view of monolayer 3$\times$3$\times$1 1T TMDs. The larger balls represent M, while the smaller ones represent X in MX$_2$.}
    \label{fig:fig4}
\end{figure}
Next, we have extended the application of our methods to monolayers of transition metal dichalcogenides (TMDs). Two forms of structure of TMD monolayers, namely, monolayer-hexagonal (1H) and monolayer-trigonal (1T), are considered, as shown in Figure \ref{fig:fig4}. We have assessed a range of meta-GGA approximations with increasing nonlocality for the band gaps of monolayers of TMDs. Beside the available experimental band gap values, the available results from the screened hybrid functional HSE06 and the many-body perturbation G$_0$W$_0$ are also included for comparison. Table 1 shows the band gap results from different methods for MoS$_2$, MoSe$_2$, WS$_2$, and WSe$_2$ 1H monolayers. As can be seen, PBE generally underestimates the band gaps of TMD monolayers, due to the same reason mentioned for the case of bulk solids. Unlike for bulk solids, SCAN remains almost the same quality as PBE for the 1H monolayers for band gaps, as shown in Table 1.  
A noticeable difference is however, observed for the TASK meta-GGA. TASK improves the values of band gaps upon PBE and SCAN for both 1H and 1T monolayers. mTASK predicts a better band gap results than TASK compared to PBE and SCAN. In mTASK, the delicate control is better realized with the increased slope of  $\frac{\partial F_X}{\partial\alpha}$ for the two-dimensional layers, leading to an improved description of band gaps. HSE06 yields a noticeable overestimation for the band gaps of MoS$_2$ and MoSe$_2$. HSE06 is based on PBE0 for the short-range. TASK and mTASK both have a more delicate set of ingredients than the short-range PBE component of HSE06 so that these meta-GGA’s can be fine tuned to find the band gap.

\begin{table}[h!]
\caption {Band gaps (eV) of 1H TMD monolayers from PBE, SCAN, TASK, and mTASK density functional approximations in comparison with those from HSE06 and experiments.}
\begin{tabular}{ccccccc}
\textbf{TMDs}  & \textbf{PBE} & \textbf{SCAN} & \textbf{TASK} & \textbf{mTASK} & \textbf{HSE} & \textbf{Expt.} \\
\hline
\hline
\textbf{MoS$_{2}$} & 1.66  & 1.64  & 1.79 & 1.80  & 2.02 \cite{kang2013band} & 1.88 \cite{mak2010atomically}         \\
\textbf{MoSe$_{2}$}  & 1.44  & 1.56   & 1.62 & 1.61   & 1.72 \cite{kang2013band}  &   1.57 \cite{lu2014large}  \\
\textbf{WS$_{2}$}& 1.81  & 1.79  & 1.94 & 1.98  & 1.98 \cite{kang2013band}  & 2.01 \cite{tongay2014tuning}          \\
\textbf{WSe$_{2}$}  & 1.53   & 1.53  & 1.66  &1.69  & 1.63 \cite{kang2013band} & 1.67 \cite{tonndorf2013photoluminescence}   \\
\hline
\hline
\end{tabular}
\end{table}

1T monolayers have not been synthetized, therefore no experimental band gaps are available for these materials. Therefore, the hybrid HSE06 and G$_0$W$_0$@PBE results have been utilized as references for comparison. Table 2 displays the band gaps of HfS$_2$, HfSe$_2$, ZrS$_2$ and ZrSe$_2$ monolayers.
The conclusion for 1T monolayers is different from the one of 1H monolayers. The SCAN meta-GGA opens the band gaps significantly compared to PBE. The TASK meta-GGA adds an even much larger amount of nonlocality to SCAN, and opens the band gaps even further. mTASK provides more refinement in slightly increased band gaps from TASK. mTASK is not only matching HSE06’s accuracy, but predicts even better band gaps for 1T TMD monolayers.

\begin{table}[h!]
\caption{Band gaps (eV) of 1T TMD monolayers from PBE, SCAN, TASK, mTASK  density functional approximations  in comparison with the available results from HSE06 and G0W0@PBE.}
\begin{tabular}{ccccccc}

\textbf{TMDs} & \textbf{PBE} & \textbf{SCAN} & \textbf{TASK} & \textbf{mTASK} & \textbf{HSE06}\cite{zhao2017elastic} & \textbf{G0W0}\cite{ZH13} \\
\hline
\hline
\textbf{HfS$_2$}  & 1.30 & 1.61  & 2.22  & 2.35  & 2.40 & 2.45  \\
 \textbf{HfSe$_2$}  & 0.63   & 0.93   & 1.51  & 1.59   & 1.32 & 1.39 \\
  \textbf{ZrS$_2$}  & 1.17  & 1.53  & 2.00  & 2.17   & 2.16 & 2.56   \\
  \textbf{ZrSe$_2$}  & 0.48 & 0.82 & 1.26  & 1.38  & 1.07 & 1.54 \\
\hline
\hline
\end{tabular}
\end{table}

\subsection{Bulk and edge-state band gaps of flat and bent nanoribbons}
\begin{figure}[h!]
    \centering
    \includegraphics[scale=0.4]{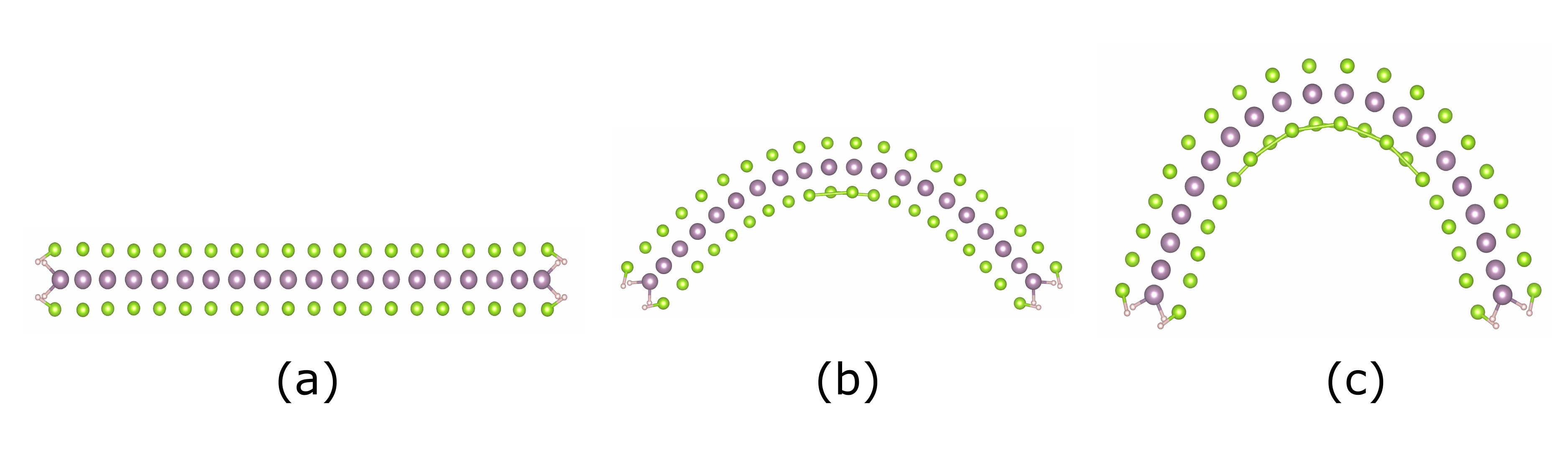}
    \caption{The cross-section views of the 1H nanoribbons with     different bending curvatures (1/R). (a) flat ribbon, R =      $\infty$, (b) R = 15 \AA, (c) R = 10 \AA. The periodical      direction of the nanoribbon is perpendicular to the page of   paper. All nanoribbons are hydrogen passivated on the two     ribbon edges and their geometrical structures are fully       relaxed as shown in (a) through (c). The blue ball            represents the M atom, the green ball represents X in         MX$_2$, and the white ball represents the H atom.}
    \label{fig:fig5}
\end{figure}
Mechanical bending can effectively control the conductivity of 2D semiconducting nanoribbons, since bending can induce non-uniform strain in nanoribbons. The magnitude of the induced strain by bending can be much larger than that of uniaxial strain \cite{YRP16,NYYR19}. Beside the delocalized states, the localized edge states gain a high relevance in the electronic structures of nanoribbon systems, while in monolayer systems of the same constituents, the edge states are usually absent. 
Some of the authors found earlier that the donor-like in-gap edge-states of armchair MoS$_2$ nanoribbon and their associated harmful Fermi-level pinning can be removed to some extent by bending \cite{YRP16}.

The nature of edge states differs significantly in 1T and 1H nanoribbons \cite{NYYR19}. The 1T nanoribbons only have the edge states below the Fermi level while both the edge states above and below the Fermi level are present in the 1H nanoribbons \cite{NYYR19}. The position of these edges states in the band structure can be critical for water splitting reactions \cite{CAA07,ZH13}, and any accurate simulation of these band gaps can be critical to guide experiments.

\begin{figure}[h!]
    \centering
    \includegraphics[scale=0.4]{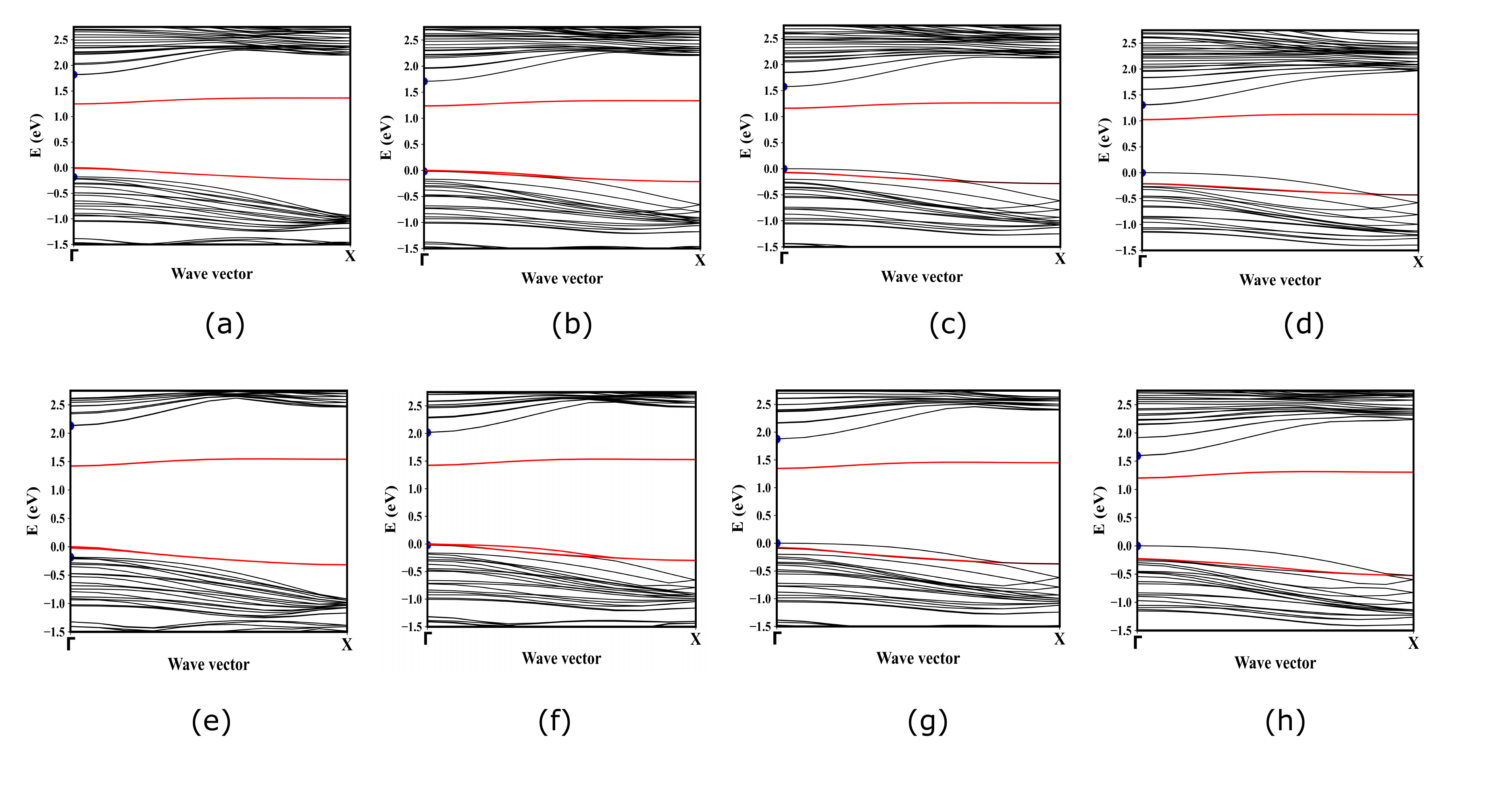}
    \caption{Band structures of armchair 1H MoS$_2$ nanoribbon obtained from the mTASK density functional at different radii of curvature (a) flat R = $\infty$ (b) R = 15 \AA\ (c) R = 12.5 \AA\ (d) R = 10 \AA\ and the hybrid HSE06 functional at different radii of curvature (e) flat R = $\infty$ (f) R = 15 \AA\ (g) R = 12.5 \AA\ (h) R = 10 \AA. The edge states are indicted by solid red lines.}
    \label{fig:fig6}
\end{figure}

\begin{figure}[h!]
    \centering
    \includegraphics[scale=0.6]{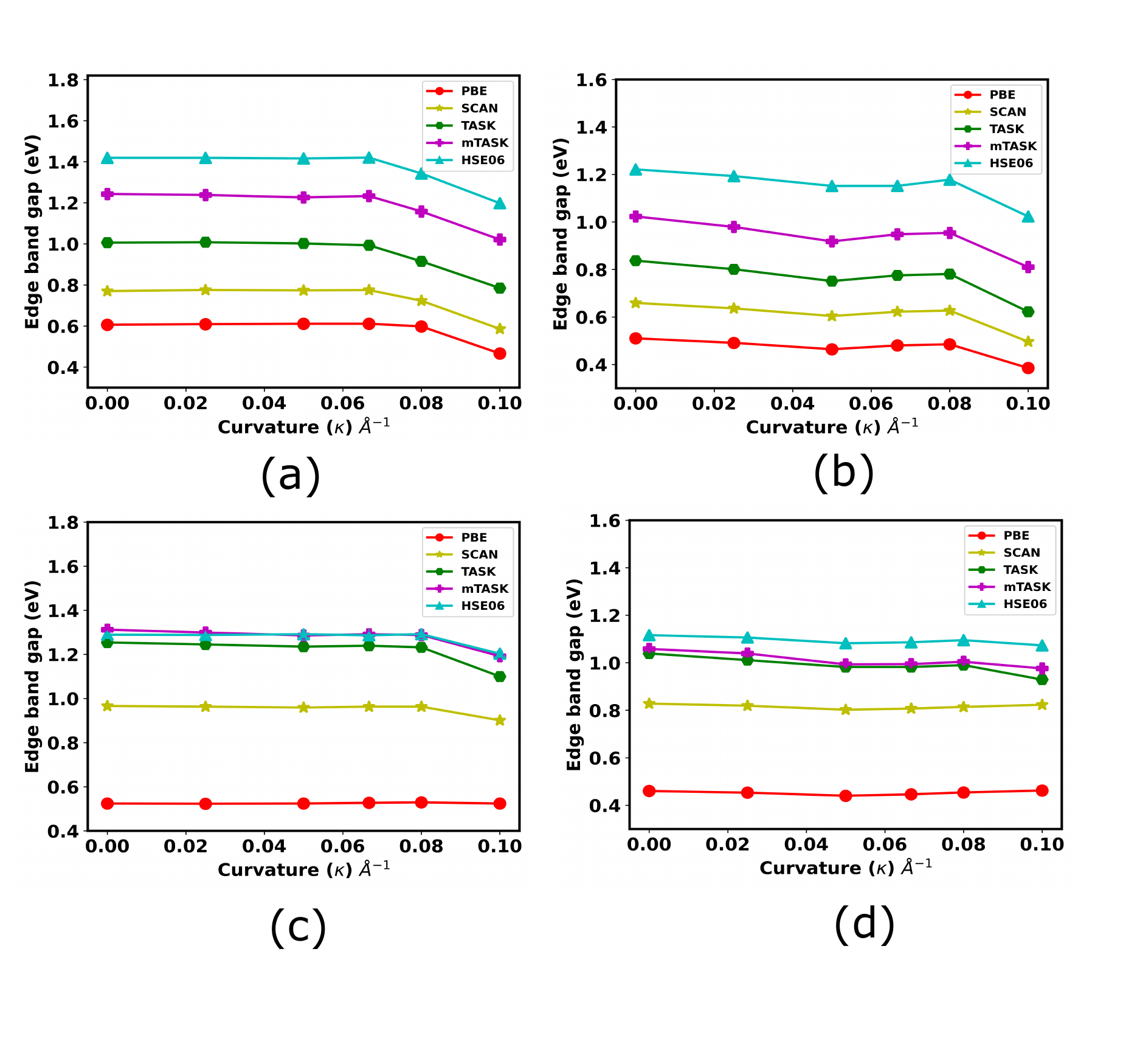}
    \caption{The band gap of the edge states of 1H TMDs nanoribbons (a) MoS$_2$. (b) MoSe$_2$, (c) WS$_2$, and (d) WSe$_2$ at different bending curvatures with different density functional approximations. Bending curvature $\kappa = \frac{1}{R}$.}
    \label{fig:fig7}
\end{figure}

1H semiconductors have a different band structure. 
The band structures of bent 1H TMDs nanoribbons evaluated with mTASK are similar to those from the hybrid HSE06. Figure \ref{fig:fig6} shows the band structures of armchair 1H MoS$_2$ nanoribbon obtained from mTASK and HSE06 at different radii of curvature. As can be seen, by increasing the bending curvature from flat (R = $\infty$) to R = 15 \AA, the upper edge band shifts towards the conduction band continuum slightly, and the lower edge band begins to touch (or merge) with the valence band continuum at R = 15 \AA. Further increase in the curvature from R =15 \AA \space to R = 10 \AA \space makes the upper edge state shift downward slightly. Also note that the minimum of the conduction band continuum is shifted downwards. This makes an overall effect that the upper edge band is gradually closer to the conduction band continuum with an increase in curvature. However, the upper edge bands never merge into the conduction band continuum. The position of the upper edge band is comparable to HSE06 level calculation for mTASK where PBE,  SCAN, and even  TASK fail to predict the correct position of the upper edge band.\\
As can be seen in figure \ref{fig:fig6}, the lower edge state begins to merge into the bulk valence band continuum at R=15  \AA, and it is eventually completely embedded at R = 10 \AA. When the lower edge states merge into the valence band continuum, the TMD nanoribbons become n-type semiconductors, given only the upper edge bands are in the gap region and above the Fermi level (or the chemical potential level). It is important to accurately predict the positions of those edge states when the harmful Fermi-level pinning needs to be evaluated for contact engineering, in applications of the semiconducting nanoribbons. Furthermore, in photocatalytic applications, such as water splitting, it is crucial that the shift of the edge states under bending is in favor of efficiency, and preserves the photocatalytic properties of the semiconductors. This feature appears in MoS$_2$ and WS$_2$ where it is highly preferred to keep the band edges in the positions where they straddle the water redox potentials \cite{ZH13}.\\
Figure \ref{fig:fig7} shows the edge band gaps of four 1H TMD nanoribbons (MoS$_2$, MoSe$_2$, WS$_2$, and WSe$_2$) at different bending curvatures, calculated from PBE, SCAN, TASK, mTASK and hybrid HSE06. Here, we define the band gap involving only the delocalized-states as the ``non-edge" or ``bulk" band gap, otherwise, it is simply referred as ``edge band gap" or simply ``band gap". As shown in figure \ref{fig:fig7} (c) and (d), the more localized edge band gaps are very accurately captured by TASK and mTASK for WS$_2$, and WSe$_2$. TASK and mTASK produce much better band gaps of MoS$_2$ and MoSe$_2$ than PBE and SCAN, when compared with a reference of HSE06. 
 The accuracy and the computational gain are apparent from mTASK, as long as sizeable nonlocality is found even at the meta-GGA level, with a reasonable agreement with HSE06. Table 3 summarizes the bulk band gaps for all 1H nanoribbons with all methods. The more localized edge-state band gaps of Figure \ref{fig:fig7} are very accurately captured by TASK and mTASK for WS$_2$, and WSe$_2$, but the agreement becomes less accurate for the more delocalized band gaps.\\

\begin{figure}
    \centering
    \includegraphics[scale=0.4]{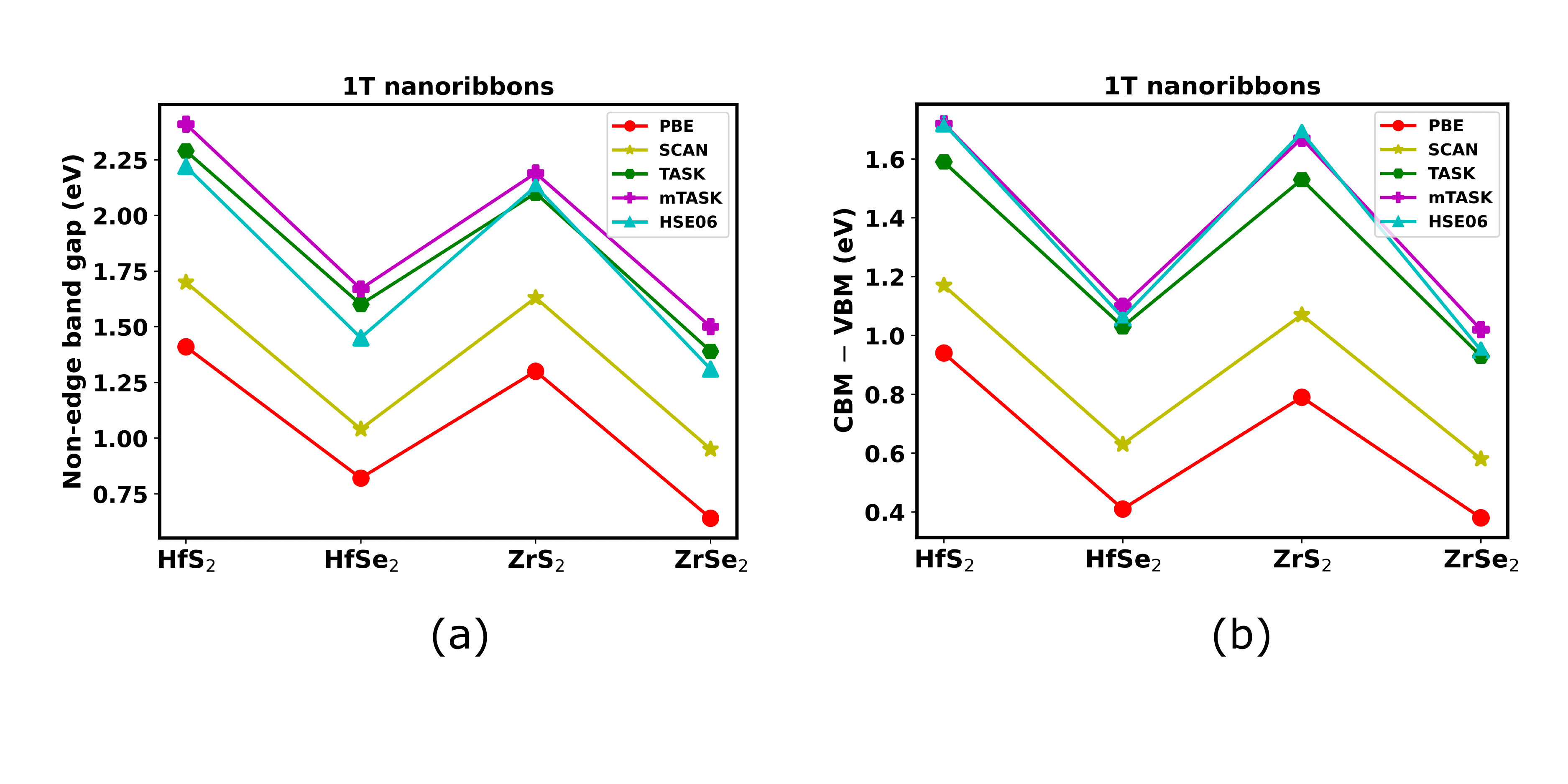}
    \caption{Band gaps of four 1T flat nanoribbons with hydrogen passivation, calculated from different density functional approximations. (a) displays the non-edge band gaps, which are measured as the gap between the maximum of the valence band continuum (excluding edge bands) and the minimum of the conduction band continuum (excluding edge bands), at the $\Gamma$ point in the band structure. (b) displays the edge band gaps, which are measured as the gap between the maximum of valence bands (including edge bands) and the minimum of conduction bands (including edge bands), at the $\Gamma$ point in the band structure. }
    \label{fig:fig8}
\end{figure}

\begin{table}[h!]
\caption{Delocalized (non-edge) band gaps of armchair 1H TMD nanoribbons at different radii of curvature (R in \AA) from different density functional approximations.}
\begin{tabular}{cllllll}
\hline
\hline
\textbf{MX$_2$}            & \multicolumn{1}{c}{\textbf{R}} & \multicolumn{1}{c}{\textbf{PBE}} & \multicolumn{1}{c}{\textbf{SCAN}} & \multicolumn{1}{c}{\textbf{TASK}} & \multicolumn{1}{c}{\textbf{mTASK}} & \multicolumn{1}{c}{\textbf{HSE06}} \\
\hline
\hline
\multirow{6}{*}{\textbf{MoS$_2$}}  &  $\infty$   & 1.76 & 1.86  & 1.91 & 2.00  & 2.32  \\
& 40 & 1.79 & 1.90 & 1.95 & 2.03 & 2.34\\
& 20 & 1.69 & 1.79  & 1.84  & 1.92 & 2.24 \\
& 15 & 1.50 & 1.59 & 1.66 & 1.73 & 2.03  \\
& 12.5 & 1.36 & 1.45 & 1.50 & 1.58 & 1.88  \\
& 10 & 1.14 & 1.19  & 1.25 & 1.31 & 1.77 \\
                                \hline
\multirow{6}{*}{\textbf{MoSe$_2$}} & $\infty$  & 1.57  & 1.70  & 1.77  & 1.76 & 2.05 \\
 & 40  & 1.60  & 1.72  & 1.80  & 1.79   & 2.08   \\
 & 20   & 1.55   & 1.67  & 1.74   & 1.74   & 2.03   \\
 & 15  & 1.42  & 1.53  & 1.57  & 1.62  & 1.90  \\
 & 12.5  & 1.14  & 1.27   & 1.33   & 1.37   & 1.66  \\
  & 10  & 0.90 & 0.99  & 1.02 & 1.07   & 1.34    \\
                                \hline
\multirow{6}{*}{\textbf{WS$_2$}}   & $\infty$  & 2.00  & 2.03   & 2.16   & 2.23  & 2.54  \\
 & 40 & 2.03  & 2.05  & 2.20  & 2.25 & 2.56 \\
& 20 & 1.94  & 2.01  & 2.09  & 2.18  & 2.49  \\
  & 15  & 1.75  & 1.82  & 1.90  & 1.96   & 2.39    \\
 & 12.5  & 1.48  & 1.59  & 1.74  & 1.78  & 2.06 \\
& 10 & 1.34 & 1.40  & 1.46 & 1.54 & 1.83  \\
                                \hline
\multirow{6}{*}{\textbf{WSe$_2$}}  &  $\infty$  & 1.73   & 1.76  & 1.89  & 1.92  & 2.22 \\
  & 40 & 1.74 & 1.77 &   1.90   & 1.94 & 2.24 \\
  & 20  & 1.70 & 1.73  & 1.86  & 1.89  & 2.19   \\
 & 15 & 1.60 & 1.63  & 1.75 & 1.78  & 2.08  \\
 & 12.5  & 1.21  & 1.33  & 1.44 & 1.47  & 1.74  \\
 & 10  & 1.07 & 1.18  & 1.22 & 1.28  & 1.54 \\    \hline
                                \hline
\end{tabular}
\end{table}
Similarly, Figure \ref{fig:fig8} plots the band gaps calculated with four semilocal density functional approximations (PBE, SCAN, TASK, and mTASK) and hybrid HSE06 for four 1T flat nanoribbons, namely, HfS$_2$, HfSe$_2$, ZrS$_2$ and ZrSe$_2$. It is noticeable that TASK and mTASK open the band gaps of both non-edge and edge-states more than SCAN or HSE06. This is consistent with the results of the 1T TMD monolayers shown in Table 2. 

\pagebreak
\section{Conclusions}
 We have systematically assessed the fundamental band gaps of semiconductors starting from bulk crystals to 1D nanoribbons. For low-dimensional materials, we have modified the TASK meta-GGA within the generalized Kohn-Sham framework \cite{YPS16} in order to increase the nonlocality in the exchange-correlation potential. The mTASK greatly improves the band gaps and is in close agreement to the experimental or the hybrid level HSE06 density functional for 2D single-layer and nanoribbon systems. This improvement for the band gap is related to the slope of the exchange enhancement factor with respect to the dimensionless orbital indicator ingredient. The physics of choosing the nonlocality in mTASK is analogous with choosing the exchange mixing factor in hybrid functionals. The global or screened hybrids PBE0 and HSE06 apply the exchange mixing globally while the TASK and mTASK meta-GGA’s change their $\frac{\partial F_X}{\partial\alpha}$ semilocally. The limits of the global exchange mixing factor are the well known values of zero and one, but except its upper bound of zero, no in-principle lower limit can be determined for $\frac{\partial F_X}{\partial\alpha}$.  Based on its orbital-dependence, $\alpha$ of meta-GGA’s is more than a local ingredient, and meta-GGA’s are more than just semilocal density functionals.
 TASK was already found to yield excellent band gaps of bulk solids. Low-dimensional materials such as monolayers and nanoribbons are known to exhibit strongly reduced screening by their reduced dimensionality. First we have investigated to what extent the accuracy of the TASK meta-GGA is transferable to 2D materials. The overall performance of mTASK for the band gaps of low-dimensional materials is better than that of TASK because of the increased nonlocality within the generalized Kohn-Sham scheme. The agreement of the band gaps and the band structures for the 2D single layers and nanoribbons quantitatively and qualitatively is the solid evidence for the accuracy of the mTASK functional for band gaps. Furthermore, the band structure obtained from mTASK for flat and bent nanoribbons is comparable to the accuracy of HSE06 calculations. The mTASK meta-GGA accurately predicts the position of edge-state bands for flat and bent nanoribbons compared to the hybrid HSE06 functional.\\

As pointed out, TASK and mTASK have a great potential for optical response properties. The same prediction applies to electronic structure simulations with structures that are different from their equilibrium. The optical response of strained and bent low-dimensional systems within time dependent density functional theory (TDDFT) is of great interest for device applications.  TDDFT is an affordable tool to evaluate optical properties. Although this statement is true in principle, the application of TDDFT is not so straightforward at this time, especially for strained or bent 2D materials. When TDDFT is applied to a bulk crystal, the optical spectrum can be evaluated on a scissor-operator-corrected DFT band gap \cite{LA89,DG93}. This approach is simpler than applying the more expensive GW approximation, but the scissor correction is often taken from experiments and the correction is very unlikely to be available for the bent structures at different bending curvatures. In bulk crystals, GW yields a nearly constant shift to the fundamental gap of semilocal DFT even for strained structures, but this is not necessarily true any more for low-dimensional systems such as nanoribbons. This difference is a consequence of enhanced many-body effects present in low-dimensional materials \cite{WSS19}. HSE06 can be feasible for the bent nanoribbons, but the HSE06 reference makes TDDFT more expensive. TASK or mTASK meta-GGA’s, are promising references for TDDFT that can make TDDFT feasible for bent or strained structures where the scissor correction may not be available. These methods can then become practical for the evaluation of exciton binding energies on strained nanoribbons.

\section{Acknowledgment}
The authors thank Prof. John P. Perdew for useful comments on the manuscript.
 This material is based upon work supported by the U.S. Department of Energy, Office of Science, Office of Basic Energy Sciences, under Award Number DE-SC0021263. The calculations were carried out on HPC resources supported in part by the National Science Foundation through major research instrumentation grant number 1625061 and by the US Army Research Laboratory under contract number W911NF-16-2-0189.

\clearpage

\begin{center}
\textbf{Supplementary Materials}
\end{center}

For slowly varying densities with s $\approx$ 0, and $\alpha \approx$ 1, one can get the condition for $\mu_\alpha$
 from the fourth-order gradient expansions (GE4) as:
 \begin{equation}
     \mu_\alpha^\pm=-\frac{97+3h_X^0\pm\sqrt{9\left(h_X^0\right)^2+74166h_X^0-64175}}{1200}
 \end{equation}
\noindent For $h_X^0=$ 1.29 , $\mu_\alpha^+ \approx - 0.231993$ and $\mu_\alpha^- \approx 0.063877$.
The more negative solution is chosen in mTASK (similar to TASK) to get sizeable nonlocality,
$\mu_\alpha^+ \approx -0.231993$.
The interpolation functions $h_X^1\left(s\right)$  and $f_X\left(\alpha\right)$ are written as Chebyshev expansions. 
\begin{equation}
    h_X^1\left(s\right)=\sum_{\nu=0}^{2}a_\nu R_\nu\left(s^2\right),\ \ f_X\left(\alpha\right)=\sum_{\nu=0}^{4}b_\nu R_\nu\left(\alpha\right)
\end{equation}

For mTASK, we choose, $ f_X\left(\alpha\rightarrow\infty\right)=-3.5$.
We found that the eight coefficients are $a_0\approx 0.924374$, $a_1\approx -0.09276847$, $a_2\approx -0.017143$, $b_0\approx\ -0.639572$, $b_1\approx -2.087488$, $b_2= -0.625$, $b_3\approx -0.162512$, and $b_4\approx 0.014572$.

\begin{figure}[h!]
    \renewcommand\thefigure{S1}
    \centering
    \includegraphics[scale=0.6]{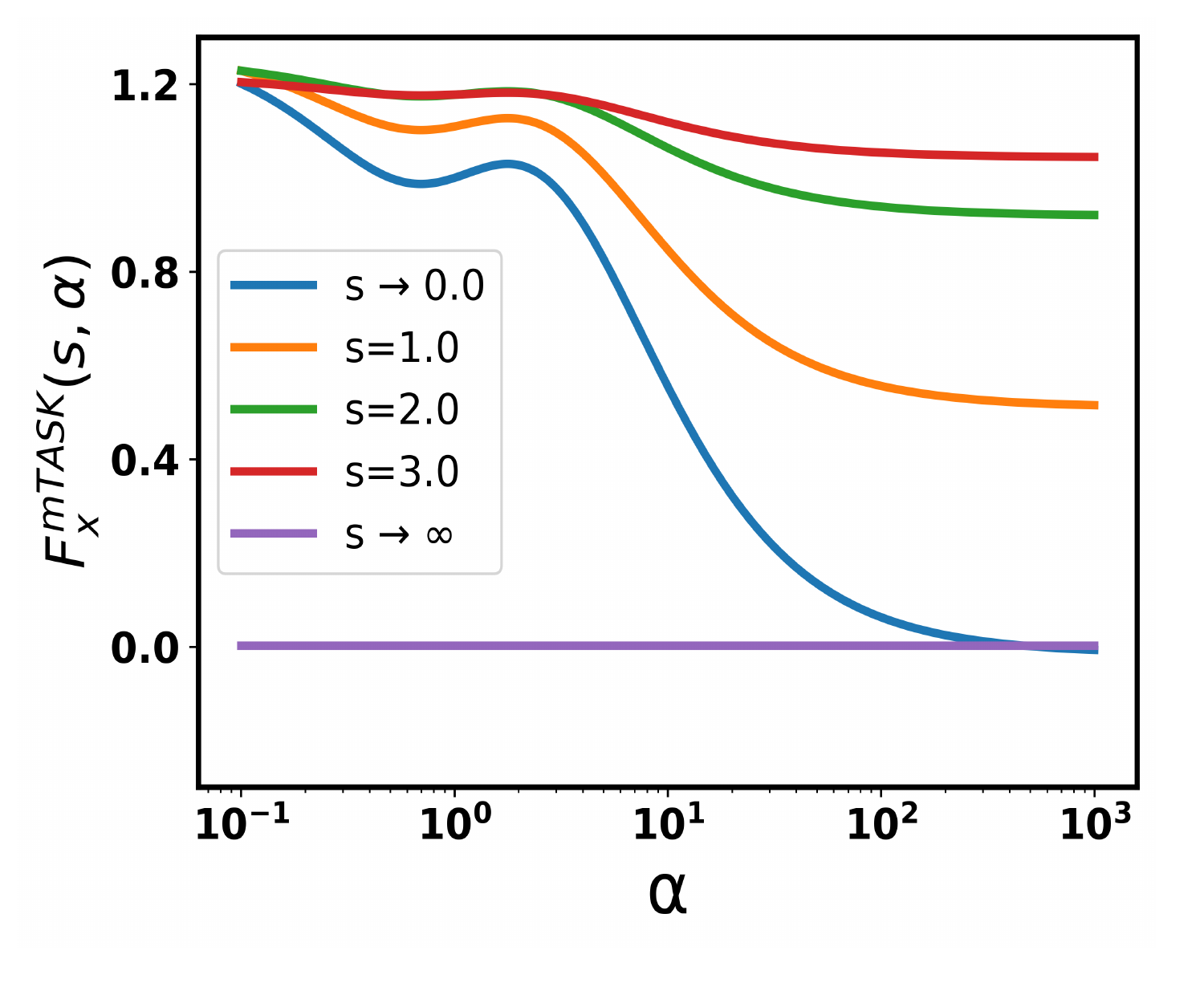}
    \caption{ Exchange enhancement factor of mTASK for more positive value of $\mu_\alpha^-=0.0638768$ as a function of $\alpha$ for fixed value of s. }
    \label{fig:my_label}
\end{figure}

The mTASK enhancement factor is not smooth for $\mu_\alpha^-=0.0638768$,  and is not a monotonically decreasing function of $\alpha$. This condition is associated with a convergence issue. We had convergence problems while testing for band gaps of bulk solids.

\begin{table}[h!]
 \renewcommand\thetable{S1}
\caption{The width of flat nanoribbons used in the calculations (\AA). The nanoribbons considered here have 20 and 12 MX$_2$ formula units in super-cell respectively, for the 1H and 1T phases. }
\begin{tabular}{cccll}
\multicolumn{1}{l}{\textbf{Phase}}&\textbf{MX$_2$}& \textbf{Width (\AA)} \\
\hline
\hline
\multirow{4}{*}{\textbf{1H}}       &  \textbf{MoS$_2$}       & 30.43  \\
                          &  \textbf{MoSe$_2$}    & 31.74   \\
                          &  \textbf{WS$_2$}       & 30.44   \\
                          &   \textbf{WSe$_2$}      & 31.74   \\
\hline
\multirow{4}{*}{\textbf{1T}}       &   \textbf{HfS$_2$}     & 36.57   \\
                          & \textbf{HfSe$_2$}      & 37.88   \\
                          &  \textbf{ZrS$_2$}      & 37.05   \\
                          & \textbf{ZrSe$_2$}      & 38.19  \\
\hline
\end{tabular}
\end{table}

\begin{figure}[h!]
 \renewcommand\thefigure{S2}
    \centering
    \includegraphics[scale=0.45]{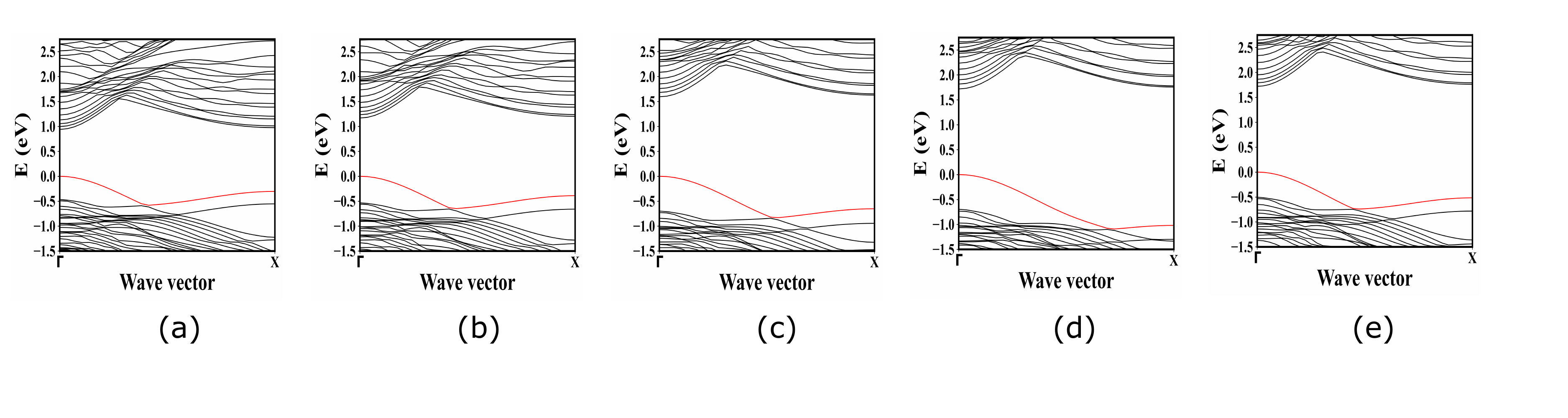}
    \caption{The band gap of the non-edge states of 1H TMDs nanoribbons (a) MoS$_2$. (b) MoSe$_2$, (c) WS$_2$, and (d) WSe$_2$ at different bending curvatures with different density functional approximations. }
    \label{fig:my_label}
\end{figure}

\begin{figure}[h!]
 \renewcommand\thefigure{S3}
    \centering
    \includegraphics[scale=0.45]{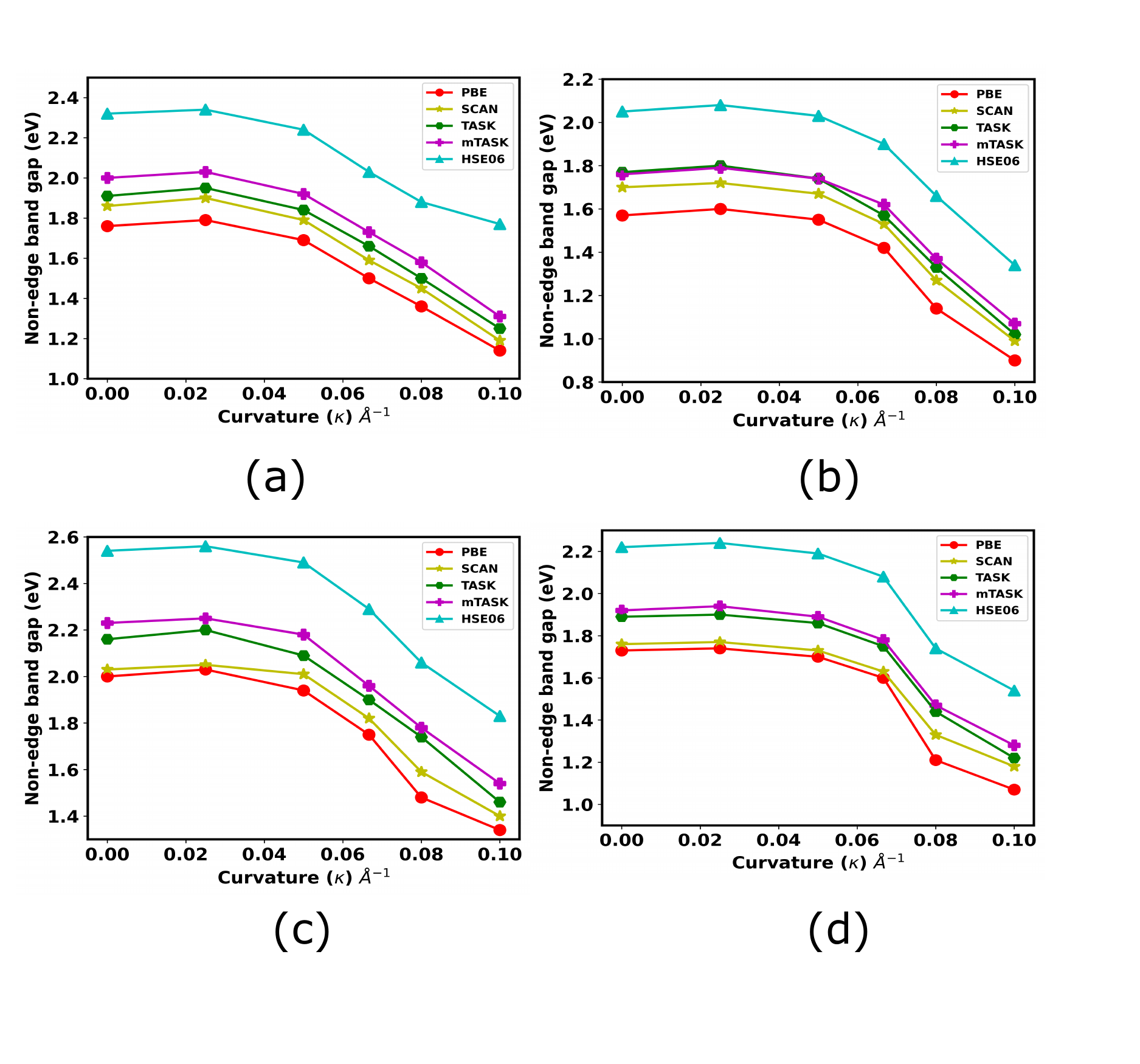}
    \caption{ Band structures of flat 1T HfS$_{2}$ nanoribbon in eVs from (a) PBE, (b) SCAN, (c) TASK, (d) mTASK, and (e) HSE06 density functional approximations. The pattern of the band structures is similar for all 1T nanoribbons that we have reported in this paper. }
    \label{fig:my_label}
\end{figure}

\end{document}